\documentclass[preprint,10pt]{elsarticle}
\usepackage{fullpage}

\usepackage{graphicx}
\usepackage[utf8]{inputenc}
\usepackage[T1]{fontenc}

\usepackage{amssymb}
\usepackage{amsthm}
\usepackage{url}
\usepackage{pdflscape}
\usepackage{afterpage}
\usepackage{capt-of}
\usepackage{array}

\usepackage{hyperref}

\newcolumntype{P}[1]{>{\centering\arraybackslash}p{#1}}
\usepackage{enumitem}
\usepackage{amsmath}
\usepackage{algorithm}
\usepackage[noend]{algpseudocode}
\let\oldReturn\Return
\renewcommand{\Return}{\State\oldReturn}

\makeatletter
\def\BState{\State\hskip-\ALG@thistlm}
\makeatother

\makeatletter
\def\ps@pprintTitle{%
   \let\@oddhead\@empty
   \let\@evenhead\@empty
   \def\@oddfoot{\reset@font\hfil\thepage\hfil}
   \let\@evenfoot\@oddfoot
}
\makeatother

\journal{}

\begin{document}

\begin{frontmatter}

\title{A review of two decades of correlations, hierarchies, networks and clustering in financial markets}

\author[e1]{Gautier Marti}
\ead{gautier.marti@polytechnique.edu}
\author[e3]{Frank Nielsen}
\author[e4]{Miko{\l}aj Bi\'nkowski}
\author[e2]{Philippe Donnat}

\address[e1]{HKML Research Limited}
\address[e2]{Hellebore Capital Ltd.}
\address[e3]{Ecole Polytechnique}
\address[e4]{Imperial College London}

\begin{abstract}
We review the state of the art of clustering financial time series and the study of their correlations alongside other interaction networks.
The aim of this review is to gather in one place the relevant material from different fields, e.g. machine learning, information geometry, econophysics, statistical physics, econometrics, behavioral finance.
We hope it will help researchers to use more effectively this alternative modeling of the financial time series. Decision makers and quantitative researchers may also be able to leverage its insights.
Finally, we also hope that this review will form the basis of an open toolbox to study correlations, hierarchies, networks and clustering in financial markets.
\end{abstract}

\begin{keyword}
Financial time series \sep Cluster analysis \sep Correlation analysis \sep Complex networks \sep Econophysics \sep Alternative data
\end{keyword}

\end{frontmatter}


\textbf{Disclaimer:} 
Views and opinions expressed are those of the authors and do not necessarily represent official positions of their respective companies.

\section{Introduction}

Since the seminal paper of Mantegna in 1999, many works have followed, and in many directions (e.g. statistical methodology, fundamental understanding of markets, risk, portfolio optimization, trading strategies, alphas), over the last two decades. We felt the need to track the developments and organize them in this present review.

\section{The standard and widely adopted methodology}

The methodology which is widely adopted in the literature stems from Mantegna's seminal paper \cite{mantegna1999hierarchical} (cited more than 1550 times as of 2019) and chapter 13 of the book \cite{mantegna1999introduction} (cited more than 4400 times as of 2019) published in 1999. We describe it below:

\begin{itemize}[noitemsep]
\item Let $N$ be the number of assets.
\item Let $P_i(t)$ be the price at time $t$ of asset $i$, $1 \leq i \leq N$.
\item Let $r_i(t)$ be the log-return at time $t$ of asset $i$: $$r_i(t) = \log P_i(t) - \log P_i(t-1).$$
\item For each pair $i,j$ of assets, compute their correlation: $$\rho_{ij} = \frac{\langle r_i r_j \rangle - \langle r_i \rangle \langle r_j \rangle}{\sqrt{\left( \langle r_i^2 \rangle - \langle r_i \rangle^2 \right) \left( \langle r_j^2 \rangle - \langle r_j \rangle^2 \right)}}.$$
\item Convert the correlation coefficients $\rho_{ij}$ into distances:
$$d_{ij} = \sqrt{2(1-\rho_{ij})}.$$ 
\item From all the distances $d_{ij}$, compute a minimum spanning tree (MST) using, for example, Algorithm~\ref{build_mst}:
\begin{quote}

\begin{algorithm}
\caption{Kruskal's algorithm}\label{build_mst}
\begin{algorithmic}[1]
\Procedure{BuildMST}{$\{ d_{ij} \}_{1 \leq i,j \leq N}$}
\State $\triangleright$ Start with a fully disconnected graph $G = (V,E)$
\State $E \gets \emptyset$
\State $V \gets \{ i \}_{1 \leq i \leq N}$
\State $\triangleright$ Try to add edges by increasing distances
\For{$(i,j) \in V^2$\text{ ordered by increasing }$d_{ij}$}
\State $\triangleright$ Verify that $i$ and $j$ are not already connected by a path
\If {\text{not} $\textit{connected}(i,j)$}
\State $\triangleright$ Add the edge $(i,j)$ to connect $i$ and $j$
\State $E \gets E \cup \{(i,j)\}$
\EndIf
\EndFor
\State $\triangleright$ $G$ is the resulting MST
\Return{$G = (V,E)$}\
\EndProcedure
\end{algorithmic}
\end{algorithm}

\end{quote}
Several other algorithms are available to build the MST \cite{huang2009comparison}.
\end{itemize}

The methodology described above builds a tree, i.e. a connected graph with $N-1$ edges and no loop.
This tree is unique as soon as all distances $d_{ij}$ are different.
The resulting MST also provides a unique indexed hierarchy \cite{mantegna1999introduction} which corresponds to the one given by the dendrogram obtained using the Single Linkage Clustering Algorithm.


\section{Methodological concerns and extensions}

Several papers have highlighted the potential shortcomings of the original methodology. We found many references which raise concerns, and suggest alternative methods. However, to this day, it does not seem that any alternatives have gained a broad enough acceptance so that they are systematically adopted for empirical studies of market correlations.

\subsection{Concerns about the standard methodology}

We list below the concerns that have been raised about the standard methodology during the last 20 years:
\begin{itemize}[noitemsep]
\item The clusters obtained from the MST (or equivalently, the Single Linkage Clustering Algorithm (SLCA)) are known to be \textbf{unstable} (small perturbations of the input data may cause big differences in the resulting clusters) \cite{DBLP:conf/icmla/MartiVDN15}.
\item The clustering instability may be partly due to the algorithm (MST/Single Linkage are known for the \textbf{chaining phenomenon} \cite{carlsson2010characterization}).
\item The clustering instability may be partly due to the correlation coefficient (\textbf{Pearson linear correlation}) defining the distance which is known for being \textbf{brittle to outliers}, and, more generally, not well suited to distributions other than the Gaussian ones \cite{DBLP:journals/prl/DonnatMV16}.
\item \textbf{Theoretical results} providing the statistical reliability of hierarchical trees and correlation-based networks are still \textbf{not available} \cite{tumminello2010correlation}.
\item One might expect that the higher the correlation associated to a link in a correlation-based network is, the higher the \textbf{reliability} of this link is. In \cite{tumminello2007spanning}, authors show that this is \textbf{not always observed empirically}.
\item Changes affecting specific links (and clusters) during prominent crises are of \textbf{difficult interpretation} due to the \textbf{high level of statistical uncertainty} associated with the correlation estimation \cite{song2011evolution}.
\item The standard method is somewhat \textbf{arbitrary}: A change in the method (e.g. using a different clustering algorithm or a different correlation coefficient) may yield a huge change in the clustering results \cite{lemieux2014clustering,DBLP:conf/icmla/MartiVDN15}. As a consequence, it implies huge variability in portfolio formation and perceived risk \cite{lemieux2014clustering}.
\end{itemize}

Notice that Benjamin F. King in his 1966 paper \cite{king1966market} (the first paper, to the best of our knowledge, about clustering stocks based on their historical returns; apparently unknown to Mantegna and his colleagues who reinvented a similar method) adds a final footnote which serves both as an advice and a warning for future work and applications:
\begin{quote}
One final comment on the method of analysis: this study has employed techniques that rely on finite variances and stationary processes when there is considerable doubt about the existence of these conditions. It is believed that a convincing argument has been made for acceptance of the hypothesis that a small number of factors, market and industry, are sufficient to explain the essential comovement of a large group of stock prices; it is possible, however, that more satisfactory results could be obtained by methods that are distribution free. Here we are thinking of a factor-analytic analogue to median regression and non-parametric analysis of variance, where the measure of distance is something other than expected squared deviation. In future research we would probably seriously consider investing some time in the exploration of distribution free methods.
\end{quote}

It is only but recently that researchers have started to focus on these shortcomings 
as we will observe through the research contributions detailed in the next section.

\subsection{Contributions for improving the methodology}

To alleviate some of the shortcomings mentioned in the previous section, researchers have mainly proposed alternative algorithms and enhanced distances.
Some refinements of the methodology as a whole, alongside efforts to tackle the concerns about statistical soundness, have been proposed.

\subsubsection{On algorithms}

Several alternative algorithms have been proposed to replace the minimum spanning tree and its corresponding clusters:

\begin{itemize}[noitemsep]
\item \textbf{Average Linkage Minimum Spanning Tree (ALMST)} \cite{tumminello2007spanning}; Authors introduce a spanning tree associated to the Average Linkage Clustering Algorithm (ALCA); It is designed to remedy the unwanted chaining phenomenon of MST/SLCA.
\item \textbf{Planar Maximally Filtered Graph (PMFG)} \cite{aste2005complex,tumminello2005tool} which strictly contains the Minimum Spanning Tree (MST) but encodes a larger amount of information in its internal structure.
\item \textbf{Directed Bubble Hierarchal Tree (DBHT)} \cite{song2011nested,song2012hierarchical} which is designed to extract, without parameters, the deterministic clusters from the PMFG.
\item \textbf{Triangulated Maximally Filtered Graph (TMFG)} \cite{massara2016network}; Authors introduce another filtered graph more suitable for big datasets.
\item Clustering using \textbf{Potts super-paramagnetic transitions} \cite{kullmann2000identification}; When anti-correlations occur, the model creates repulsion between the stocks which modify their clustering structure.
\item Clustering using \textbf{maximum likelihood} \cite{giada2001data,giada2002algorithms}; Authors define the likelihood of a clustering based on a simple 1-factor model, then devise parameter-free methods to find a clustering with high likelihood.
\item Clustering using \textbf{Random Matrix Theory (RMT)} \cite{plerou2000random}; Eigenvalues help to determine the number of clusters, and eigenvectors their composition.
\item \cite{PhysRevX.5.021006} proposes network-based community detection methods whose null hypothesis is consistent with RMT results on cross-correlation matrices for financial time series data, unlike existing community detection algorithms.
\item Clustering using the \textbf{$p$-median problem} \cite{kocheturov2014dynamics}; With this construction, every cluster is a star, i.e. a tree with one central node.
\end{itemize}

\subsubsection{On distances}

At the heart of clustering algorithms is the fundamental notion of distance that can be defined upon a proper representation of data. It is thus an obvious direction to explore. We list below what has been proposed in the literature so far:

\begin{itemize}[noitemsep]
\item Distances that try to quantify how one financial instrument provides information about another instrument:
\begin{itemize}
\item Distance using \textbf{Granger causality} \cite{billio2012econometric},
\item Distance using \textbf{partial correlation} \cite{kenett2010dominating},
\item Study of asynchronous, \textbf{lead-lag relationships by using mutual information} instead of Pearson's correlation coefficient \cite{fiedor2014information,rocchi2016emerging},
\item The correlation matrix is normalized using the affinity transformation:
the correlation between each pair of stocks is normalized according to the correlations of each of the two stocks with all other stocks \cite{kenett2010dynamics}.
\end{itemize}
\item Distances that aim at including non-linear relationships in the analysis:
\begin{itemize}
\item Distances using mutual information, mutual information rate, and other \textbf{information-theoretic} distances \cite{fiedor2014networks,rocchi2016emerging,baitinger2017interconnectedness,barbi2017nonlinear,goh2018inference,guo2018development},
\item The Brownian distance \cite{zhang2014systemic},
\item \textbf{Copula-based} \cite{marti2016optimal,durante2015cluster,brechmann2013hierarchical} and \textbf{tail dependence} \cite{durante2015portfolio,lohre2020hierarchical} distances.
\end{itemize}
\item Distances that aim at taking into account multivariate dependence:
\begin{itemize}
\item Each stock is represented by a bivariate time series: its returns and traded volumes \cite{brida2008multidimensional}; a distance is then applied to an \textit{ad hoc} transform of the two time series into a symbolic sequence,
\item Each stock is represented by a multivariate time series, for example the daily (high, low, open, close) \cite{lee2013multidimensional}; Authors use the \textbf{Escoufier's RV coefficient} (a multivariate extension of the Pearson's correlation coefficient).
\end{itemize}
\item A distance taking into account both the \textbf{correlation} between returns \textbf{and} their \textbf{distributions}  \cite{DBLP:journals/prl/DonnatMV16}.

\item Unlike recent studies which claim that the existence of nonlinear dependence between stock returns have effects on network characteristics, 
\cite{hartman2018nonlinearity} documents that ``most of the apparent \textbf{nonlinearity is due to univariate non-Gaussianity}. Further, strong non-stationarity in a few specific stocks may play a role. In particular, the sharp decrease of some stocks during the global financial crisis in 2008'' gives rise to apparent negative tail dependence among stocks.
When constructing unweighted stock networks, they suggest to use linear correlation ``on marginally normalized data'', that is Spearman's rank correlation. In fact, this is similar to the idea of splitting apart the dependence information from the distribution one as in \cite{DBLP:journals/prl/DonnatMV16}, where Spearman's rank correlation stems from using a Euclidean distance between the uniform margins of the underlying bivariate copula. Following previous studies, and unlike in \cite{DBLP:journals/prl/DonnatMV16}, the distribution information is discarded when constructing the network.

\end{itemize}

\subsubsection{On other methodological aspects}

Besides research contributions on algorithms and distances, other methodological aspects have been pushed further.

\begin{itemize}[noitemsep]
\item Reliability and statistical uncertainty of the methods:
\begin{itemize}
\item A \textbf{bootstrap} approach is used to estimate the statistical reliability of both hierarchical trees \cite{tumminello2007hierarchically,DBLP:conf/ijcai/MartiAND16} and correlation-based networks \cite{tumminello2007spanning,musciotto2018bootstrap},

\item \cite{marti2020corrgan} suggests the use of generative adversarial networks (\textbf{GANs}) for sampling realistic yet artificial financial correlation matrices to compare and test the robustness of results, \textbf{an alternative approach to bootstrap} methods,

\item Following \cite{marti2020corrgan}, authors of \cite{papenbrock2020matrix} propose an alternative method to generate synthetic correlations: a multi-objective evolutionary algorithm which explores the `neighborhood' of a given empirical matrix.

\item \textbf{Consistency} proof of clustering algorithms for recovering clusters defined by nested block correlation matrices; Study of empirical \textbf{convergence rates} \cite{DBLP:conf/ijcai/MartiAND16},
\item \textbf{Kullback-Leibler divergence} is used to estimate the amount of filtered information between the sample correlation matrix and the filtered one \cite{tumminello2007kullback},
\item \textbf{Cophenetic correlation} is used between the original correlation distances and the hierarchical cluster representation \cite{papenbrock2015handling},
\item Several measures between successive (in time) clusters, dendrograms, networks are used to estimate \textbf{stability} of the methods, e.g. cophenetic correlation between dendrograms in \cite{panton1976comovement}, adjusted Rand index (ARI) between clusters in \cite{DBLP:conf/icmla/MartiVDN15}, mutual information (MI) of link co-occurrence between networks in \cite{song2011evolution}.
\item In \cite{kakushadze2016statistical}, authors claim that clustering still cannot compete with ``fundamental'' industry classifications in terms of performance due to inherent out-of-sample instabilities, and thus propose to improve such given ``fundamental'' industry classification via further clustering large sub-industries at the most granular level; Similarly, Lopez de Prado in \cite{lopez2019estimation} finds that ``empirical correlation matrices are \textbf{unstable and backward-looking}, and proposes a more sophisticated approach to tackle this issue: A method which fits a correlation matrix constrained by a \textbf{hierarchical structure encoding an economic and potentially forward-looking hypothesis}.
Avellaneda, in \cite{avellaneda2020hierarchical}, builds on similar motivations to propose a hierarchical PCA yielding principal components which are constrained by prior (hierarchical) market partition (e.g. into sectors); these principal components are thus more stable and interpretable.
\end{itemize}
\item Preprocessing of the time series:
\begin{itemize}
\item \textbf{Subtract the market mode} before performing a cluster or network analysis on the returns \cite{borghesi2007emergence},
\item Encode both \textbf{rank statistics} and a \textbf{distribution histogram} of the returns into a \textbf{representative vector} \cite{DBLP:journals/prl/DonnatMV16},
\item Fit an ARMA(p,q)-FIEGARCH(1,d,1)-cDCC process (\textbf{econometric preprocessing}) to obtain dynamic correlations instead of the common approach of rolling window Pearson correlations \cite{sensoy2014dynamic}, 
\item Use a clustering of successive correlation matrices to \textbf{infer a market state} \cite{munnix2012identifying,papenbrock2015handling,rinn2015dynamics,stepanov2015stability,heckens2020uncovering}.
\end{itemize}
\item \textbf{Clustering} of the time series based on \textbf{parameters of estimated models}:
\begin{itemize}
    \item using coefficients of GARCH processes \cite{otranto2008clustering}
    \item using a Wasserstein distance between probability distributions obtained from a time series change-point model \cite{whiteley2019dynamic}
\end{itemize}{}
\item Use of \textbf{other types of networks}: threshold networks \cite{onnela2004clustering}, influence networks \cite{gao2015influence}, partial-correlation networks \cite{kenett2010dominating,kenett2012dependency}, Granger causality networks \cite{billio2012econometric,vyrost2015granger}, cointegration-based networks \cite{tu2014cointegration}, bipartite networks \cite{tumminello2011statistically}, multilayer networks \cite{baltakys2018multilayer,de2019multilayer}, Bayesian networks and other probabilistic graphical models \cite{denev2015probabilistic}, etc.
\item Understanding of the drivers of synchronous correlations using the properties of the collective stock dynamics at \textbf{shorter time scales} \cite{curme2015lead} by using \textbf{directed networks} of lagged correlations \cite{curme2015lead,curme2015emergence}.
\end{itemize}

\section{Dynamics of correlations, hierarchies, networks and clustering}

Many of the empirical studies are based on the whole period available from the data.
Some researchers have started to investigate the dynamics of the empirical correlations, and also the hierarchies, networks and clusters extracted from them (cf. \cite{onnela2003dynamics} as one of the earliest work). This dynamic setting which has the potential to track changes of the market structure is more interesting for practitioners (e.g. risk managers, traders, regulatory agencies).
This research is still in its infancy and we think its results are still hardly exploitable in practice. For instance, an interesting but difficult question is the following: Are \textbf{changes} in the correlation structure due to statistical \textbf{noise} and data artifacts \textbf{or} do they provide a \textbf{real signal}?

No predominant methodology has emerged for now but the naive one which consists in: 
\begin{itemize}
\item Computing Pearson correlations on a \textbf{rolling window of arbitrary length}, 
\item then independently computing a network or a clustering based on the rolling empirical correlation matrix.
\end{itemize}
Some promising avenue of research may be the use of temporal networks and temporal centrality measures \cite{zhao2017stock}.

Besides the shortcomings of Pearson correlation detailed above, this approach is brittle due to its strong \textbf{dependence \textit{a priori} on:}
\begin{itemize}
\item the \textbf{sampling frequency} (e.g., intraday, daily, weekly),
\begin{itemize}
\item Concerning the sampling frequency, authors in \cite{bonanno2001high} notice that at intraday frequency level some time is needed before the cluster organization emerges completely. 
According to the paper, ``the changes observed in the structure of the MST and of the hierarchical tree suggest that the intrasector correlation decreases faster than intersector correlation between pairs of stocks'' when sampling frequency increases.
In \cite{borghesi2007emergence,DBLP:conf/icmla/MartiVDN15}, authors observe that the clusters obtained using daily returns are similar to the ones obtained with weekly timescales, and even to some extent to the ones using monthly returns.
Most of the empirical studies focus on daily returns and only a few explore intraday data: \cite{bonanno2001high,johnson2005shakes,borghesi2007emergence,zhang2011will,lee2012intraday,curme2015lead}.
Working with higher frequencies (e.g. at the transaction or quote level) brings further difficulties such as coping with asynchronous data and the Epps effect \cite{epps1979comovements}.
\end{itemize}
\item the \textbf{length $T$} of the rolling window,
\begin{itemize}
\item What is the right length for the rolling window? 
No clear-cut answer has yet been proposed and, in most studies, its length is set somewhat arbitrarily.
In \cite{onnela2003dynamics}, authors posit that ``the choice of window width is a trade-off between too noisy and too smoothed data for small and large window widths, respectively'' and that they ``have explored a large scale of different values for both parameters, and the given values were found optimal''.
What are the proper criteria for setting the window length? 
The choice can be driven by the goal (e.g. time investment horizon),
by regulatory rules (e.g. computing Value-at-Risk using 1-year historical data),
by the stability of clusters \cite{DBLP:conf/icmla/MartiVDN15},
by a statistical convergence rate \cite{DBLP:conf/ijcai/MartiAND16},
by economic regimes
or by a trade-off of the preceding criteria. 
\end{itemize}
\item the \textbf{number $N$ of assets} studied.
\begin{itemize}
\item The number of considered assets has also a significant impact on the results:
the ratio $T/N$ drives the precision of the correlation matrix estimation and ultimately the clustering 
\cite{borysov2014asymptotics,DBLP:conf/ijcai/MartiAND16,bun2016rotational,bun2017cleaning}.
\end{itemize}
\end{itemize}

This dependence makes it difficult to fully understand and analyze results.
Once these `parameters', i.e. the sampling frequency, $T$, and $N$, are chosen,
one can study

\begin{itemize}[noitemsep]

\item the \textbf{dynamics of correlations}:
\begin{itemize}[noitemsep]
\item In \cite{kenett2010dynamics}, authors are using a sliding window of $T=22$ days to measure and monitor the eigenvalue entropy of the stock correlation matrices (estimated using daily returns, for $N=25$ (Tel-Aviv stock market), and $N=455$ (from S\&P 500)).
They also propose a 3D visualization to monitor the configuration of stocks using a 3D PCA.
\item \cite{meng2014systemic} notices three regime shifts during the period 1989-2011 by monitoring eigenvalues and eigenvectors of the empirical correlation matrices (estimated using quarterly
recorded prices from the US housing market; $T=60$, $N=51$, the number of US states). 
\end{itemize}

\item the \textbf{dynamics of the MST} and other hierarchical trees:
     
Using summary statistics:
\begin{itemize}[noitemsep]
\item The MST which evolves over time is monitored using summary statistics (also called topological features) \cite{onnela2003dynamic} such as the normalized tree length \cite{onnela2003dynamics}, the mean occupation layer \cite{onnela2003dynamics}, the tree half-life \cite{onnela2003dynamics}, a survival ratio of the edges \cite{onnela2002dynamic,johnson2005shakes,sensoy2014dynamic}, node degree, strength \cite{sensoy2014dynamic}, eigenvector, betweenness \cite{tang2018complexities}, closeness centrality \cite{sensoy2014dynamic}, the agglomerative coefficient \cite{matesanz2015sovereign}.
\item Using these statistics, \cite{onnela2003dynamics} notices that: 
\begin{itemize}
\item the MST strongly shrinks during a stock market crisis,
\item the optimal Markowitz portfolio lies practically at all times on the outskirts of the tree,
\item the normalized tree length and the investment diversification potential are very strongly correlated.
\end{itemize}
\item And \cite{sensoy2014dynamic} notices that in the Asia-Pacific stock market:
\begin{itemize}
\item the DST (a dynamic MST built from dynamic correlations) shrinks over time,
\item Hong Kong is found to be the key financial market,
\item the DST has a significantly increased stability in the last few years,
\item the removal of the key player has two effects: there is no clear key market any longer and the stability of the DST significantly decreases.
\end{itemize}
\item In \cite{jung2008group}, authors observe that for the Japanese and Korean stock markets, 
there is a decrease of grouping by industry categories.
\item Authors investigate the impact of the COVID-19 pandemic on sovereign bond yields amongst European countries in \cite{pang2020analysis}. They find that the average correlation between sovereign bonds within the COVID-19 period decreases from the peak observed in the 2019-2020 period; This trend is also reflected in all network filtering methods.
\end{itemize}

Using distances or similarity measures between successive dendrograms:
\begin{itemize}
\item Cophenetic correlation coefficient. In \cite{matesanz2015sovereign}, authors propose a cophenetic analysis of public debt dendrograms in the European Union ($N = 29$ countries) computed using Pearson correlation of quarterly debt-to-GDP ratios between 2000 Q1 and 2014 Q1 ($T = 57$) with a sliding window of size $w = 15$.
\end{itemize}

\item the \textbf{dynamics of clusters}:
\begin{itemize}[noitemsep]

\item The paper \cite{kocheturov2014dynamics} finds that the cluster structures are more stable during crises (using the $p$-median problem, an alternative clustering methodology).

\item Authors in \cite{johnson2005shakes} notice that there is an ``ecology of clusters'': They ``can survive for finite periods of time during which time they may evolve in some identifiable way before eventually dissipating or dying''.

\item In \cite{papenbrock2015handling}, the authors track the merging, splitting, birth, death, contraction, and growth of the clusters in time.

\end{itemize}
\end{itemize}



\section{Clusters, hierarchies, and networks based on alternative data}

In this review, we cited a number of publications computing and studying clusters, hierarchies and networks based on time series of returns and their correlations, i.e. readily available and cheap data. Another avenue of research consists of using alternative data to estimate lead-lag or contemporaneous relations between companies.

\subsection{Based on textual data}

\begin{itemize}

    \item In \cite{guo2018media}, authors suggest that media network based investors' attention is a powerful predictor of market premium: In brief, the more frequent the stocks are co-mentioned by media news, the more non-shareholders attention is triggered, and the higher probability of overvaluation for connected stocks. They create an attention index, by aggregating across all the stocks in the market on monthly basis, and find that their index can forecast the market premium with a significantly negative coefficient, a 5.97\% and a 5.80\% monthly in-sample and out-of-sample $R^2$ respectively. In addition, they show that the findings hold when controlling for alternative attention proxies, news-based predictors, fundamental information predictors and other standard factors. Their indicator consistently provides negative return forecasts for both time-series and cross-sectional portfolios. As a final test, they also provide evidence that their indicator captures investor attention by sorting cross-sectional portfolios on news co-occurrence frequencies and by checking the performance of average correlation of Google search and Bloomberg search frequencies.
    
    \item Authors in \cite{forss2017news} propose to derive a company risk indicator from news-sentiment networks. For their research, they gathered from Seeking Alpha a set of articles, which contain a body of text and the author’s self-reported sentiment regarding the targeted entities in the articles. The sentiment in the articles, can be seen as a reflection of the author’s future expectations. They build co-occurrence networks from the parsed news articles on a 
    quarterly basis, starting from Q1 2011 until the end of Q2 2016.
    These co-occurrences can cover partnerships, joint ventures, competitors, 
    and suppliers, but can also cover other type of relationships, such as for example an 
    author listing his or her opinion on current best stock picks or current worst stock picks.
    They show that the highest quarterly risk value outputted by their risk model is correlated to a higher chance of stock price decline up to 70 days after a quarterly risk measurement.

    \item Authors of \cite{ronnqvist2014text, ronnqvist2015bank} suggest to study the co-occurrence network of European banks mentioned in financial discussions to better understand systemic risk in the financial system. Unlike conventional approaches that estimate interdependence based either on non-publicly disclosed information (such as interbank asset and liability exposures) or publicly available co-movements in market data which only measure their dependence indirectly (co-movements may be driven by other factors) and may not be forward-looking (co-movements are estimated on past market data), their text-to-network process can be applied to publicly available data (such as discussions in financial forums) which may be forward-looking.
    However, authors document some of their methodology limits: ``the loose definition of co-mention  context, as an entire post, lessens reliability of relations extracted, while yielding  more  relations. Provided enough source data, the context could be narrowed down, to increase the likelihood that a relation pair is actually meaningfully related.''
    They find that centrality in the network can be explained by a set of standard variables including size (measures of total assets and total deposits).

    \item In \cite{souza2018predicting}, authors aim at predicting the probability that an edge is inserted or removed in the (correlation-based) financial network estimated on past returns.
    To do so, they use both the past past structure of this financial network and also a social network estimated from the correlations of Twitter sentiment time series provided by PsychSignal.com on the same set of stocks. They find that considering both networks considerably improves the predictability of future financial networks over a naive benchmark of persistent structure. They also find that the so-called social networks are harder to predict and much less persistent than the financial ones.
    
    \item Following \cite{hoberg2017text}, \cite{fan2019topology} builds similarity networks of US listed firms from text-based analysis of the product descriptions included in the Securities Exchange Commission (SEC) filings (Form 10-K). Authors use topological features of these networks as input to machine learning models to predict the future success or failure of the firms. When these features are combined with typical financial data such as the book-to-market ratio, measures of leverage, profitability, default risk of the firms, momentum and liquidity of their stocks, accuracy of the predictions are significantly higher.

\end{itemize}



In the next three subsections, we list a couple of papers exploring interactions of different kinds: supply chain, payments, business partnerships, financial contracts, and mutual ownership.



\subsection{Based on supply chain data}

\begin{itemize}
    \item \cite{wu2015centrality} explores ``the interactions
from the perspective of an economy-wide supply-chain network, and propose a list of network centrality
measures to capture the relative importance of each company within this network.'' Using FactSet Supply Chain Relationships database, the author estimates dynamic supplier networks and compute centrality measures for each company. From there, the author constructs a supplier central portfolio of stocks based on the top ten companies with the highest centrality in the network. The author finds that the stock performance of supplier central portfolios tends to predict the movements of the overall stock market.

\item In \cite{buraschi2012dynamic}, authors study the network structure implied by firms’ cash flows on the cross-section of expected returns. They find that firms which are more central have lower P/D ratios and higher expected returns.

\item \cite{agca2017credit} find that there is significant credit risk propagation along supply chains by analyzing credit default swap spreads.

\end{itemize}


\subsection{Based on transaction data}

\begin{itemize}
    \item In \cite{letizia2018corporate}, authors study a large proprietary dataset which consists of a network built from transactional data of the payment platform of a major European bank.
    These transactions link around 2.4 million Italian firms whose credit risk rating is known for a large fraction of them. Authors find that this network, like many financial networks, is sparse but made of a single component, scale free and verifies the small world property.
    The main contribution of the authors is to document significant correlations between local topological properties of a node (firm) and its risk. They also employ machine learning techniques to build classifiers for predicting the risk rating using as inputs network properties (e.g. degree, community) alone, i.e. no balance sheet information or any other features.
    Their classification method outperforms significantly its benchmark, i.e. random assignments.
    Another contribution is to show the existence of an  homophily  of  risk, i.e. the  tendency  of
    firms  with  similar  risk  profile  to  be  statistically  more  connected  among  themselves through payments.
    Risk is therefore not spread uniformly on the network, but rather concentrated in specific areas.
    This implies that an idiosyncratic shock on a single firm can propagate more or less quickly depending on the local network structure and the community the node belongs to.

\end{itemize}


\subsection{Based on key people}


\begin{itemize} 

    \item In \cite{ozgur2008co}, authors build a co-occurrence network of people mentioned in 21,578 Reuters news articles mostly focused on economics published in 1987: People are represented as vertices and two persons are connected if they co-occur in the same article. They observed that this network has small-world features with power-law degree  distribution. The network is disconnected and the component size distribution has
    power-law characteristics. They compare the importance of these people back in 1987 in the news co-occurrence network to the importance they have as of 2007 proxied by their Wikipedia article.
    They find medium level Spearman's rank correlations between these two measures.

\end{itemize}


\subsection{Based on investors}

\begin{itemize}

    \item Traders and investors are socially connected and have access to comparable sources of information. \cite{colla2009information} investigates social connections and information linkages, and demonstrates how they can predict patterns of trade correlation: trades generated by ``neighbor'' traders are positively correlated and trades generated by ``distant'' traders are negatively correlated.
    
    \item Using a dataset of all trades on the Istanbul Stock Exchange in 2005, \cite{ozsoylev2013investor} identifies traders with similar trading behavior as linked in an empirical investor network. They find that central investors earn higher returns and trade earlier than peripheral investors with respect to information events (e.g. earnings), consistent with the view that information diffusion among the investor population influences trading behavior and returns.
    
    \item In \cite{baltakys2019clusters}, authors study investor clusters in the first two years after the IPO filing in the Helsinki Stock Exchange by using a statistically validated network method to infer investor links based on the co-occurrences of investors’ trade timing for 69 IPO stocks. They find that the clusters of investors based on their trading are stable between new and mature stocks. Authors identify a highly persistent institutional investor cluster, which they consider as evidence about institutional herding.
    
    \item \cite{baltakys2018multilayer} investigate trading decisions for Finnish investors using multilayer networks. They find that households in the capital have high centrality in investor networks, which, under the theory of information channels in investor networks, suggests that they are well-informed investors.
    
    \item Using a  Kinetic Ising model, \cite{campajola2019unveiling} reconstructs a lead-lag network of traders client of a major dealer in the Foreign Exchange market. They identify leading players in this market who are typically leading the order flow on the 5 minutes time-scale.
    Based on this lead-lag network, authors also define a herding measure based on the observe and inferred traders' opinions. They find a causal link between herding and the liquidity in the inter-dealer market used by dealers to rebalance their inventories.

\end{itemize}

\subsection{Based on several corporate networks}

\begin{itemize}
    \item \cite{de2019multilayer} combines different types of networks (ownership links, social ties through board members, research collaborations, and stock correlations) into a single multiplex structure. Authors employs several methods to show the significance of this multilayer network. Their initial results indicate a relation between company performance and multiplex centrality.
\end{itemize}{}

\section{Financial applications}

Though many of the academic studies focus on the MST or the clusters \textit{per se},
some papers try to extend their use beyond the filtering of empirical correlation matrices.
It has been proposed to leverage them for making financial policies, optimizing portfolios,
computing alternative Value-at-Risk measures, residualizing expected returns, grouping and selecting quantitative trading alphas, etc.

\subsection{Portfolio Design}

\begin{itemize}[noitemsep]

\item \cite{onnela2003dynamics} finds that the Markowitz portfolio layer in the MST is higher than the mean layer at all times.

\item As the stocks of the minimum risk portfolio are found on the outskirts of the tree \cite{pozzi2013spread,onnela2003dynamics}, 
authors expect larger trees to have greater diversification potential.

\item In \cite{tola2008cluster,panton1976comovement}, authors compare
the Markowitz portfolios from the filtered empirical correlation matrices using the clustering approach, the RMT approach and the shrinkage approach.

\item \cite{ren2016dynamic,peralta2016network} propose to invest in different part of the MST depending on the estimated market conditions.

\item Authors show that there is no inner-mathematical relationship between the minimum variance portfolio from Markowitz theory and the portfolios designed from the minimum spanning tree \cite{huttner2018portfolio}. Empirical evidence of such relations found by previous studies is essentially a stylized fact of financial returns correlations and time series, not a general property of correlation matrices.

\item It appears that a large number of stocks are unnecessary for building an index of market change \cite{king1966market}.

\item The paper \cite{dose2005clustering} describes methods for index tracking and enhanced index tracking based on clusters of financial time series.

\item \cite{durante2015portfolio} introduces a procedure to design portfolios which are diversified in their tail behavior by selecting only a single asset in each cluster.

\item \cite{baitinger2016interconnectedness} investigates several network and hierarchy based active portfolio optimizations, and find their out-of-sample performance competitive with respect to conventional ones.

\item \cite{leon2017clustering} presents the performance of seven portfolios created using clustering analysis techniques
to sort out assets into categories and then applying classical optimization inside every
cluster to select best assets inside each asset category.

\item \cite{gava2019beyond} applies a hierarchical clustering to a set of government bond factors (investment styles such as carry, momentum, slope, convexity, reversal, real rate vs. growth), and finds that most factors benefit from a positive carry.

\item \cite{raffinot2016hierarchical,raffinot2018hierarchical,lopez2016building,de2018advances,pfitzinger2019constrained,lohre2020hierarchical} leverage hierarchical clustering to build diversified portfolios that outperform out-of-sample by refining the risk parity and the equal risk contribution methods to take into account the hierarchical correlations of assets.

\item \cite{Jain_2019} compares the Hierarchical Risk Parity (HRP) from \cite{lopez2016building} to more traditional risk-based portfolios such as inverse volatility weighted, minimum variance and maximum diversification portfolios. The main takeaway from the study is that \textbf{results strongly depend on the estimation of the covariance matrix}. If estimates are crude, then inverse volatility weighted portfolios outperform the ones obtained from the other methods as less sensitive to the covariance misspecification. HRP stands in between this simple yet robust method and the optimization-based ones. Rebalancing frequency of the portfolios also impacts the relative performance of the different methods, HRP being superior at minimizing out-of-sample portfolio variance with longer rebalancing horizons.
Note that the study only considers the HRP \cite{lopez2016building} to represent the class of machine learning-based portfolios whereas other methods are available in the literature such as the Hierarchical Clustering based Asset Allocation (HCAA) \cite{raffinot2016hierarchical} and the Hierarchical Equal Risk Contribution (HERC) \cite{raffinot2018hierarchical}.
Note also that the empirical results described in the paper were obtained for portfolios of 10 assets only. HRP, and related methods, would benefit from larger portfolios, since not inverting the covariance matrix.

\end{itemize}

\subsection{Trading Strategies}


\begin{itemize}[noitemsep]

\item In \cite{papenbrock2015handling}, they suggest that tracking the merging, splitting, birth, and death of the clusters in time could be the basis for pairs-like reversal trading strategies but with pairs corresponding to clusters.

\item One can build a simple mean-reversion statistical arbitrage strategy whereby one assumes that stocks in a given industry move together, cross-sectionally demeans stock returns within said industry, shorts stocks with positive residual returns and goes long stocks with negative residual returns \cite{kakushadze2016statistical}.

\item Earnings per share forecasts prepared on the basis of statistically grouped data (clusters) outperform forecasts made on data grouped on traditional industrial criteria as well as forecasts prepared by mechanical extrapolation techniques \cite{elton1971improved}.

\item \cite{sandhu2015market} suggests that one may design a new set of Ricci curvature network based-strategies in statistical arbitrage (e.g. for mean-reverting portfolios).

\item \cite{musmeci2016interplay} finds the existence of significant relations between past changes in the market correlation structure and future changes in the market volatility.

\item In \cite{kakushadze2016statistical}, authors suggest the use of clustering to build statistical classification of quantitative trading alphas for which there is no analog of ``fundamental'' industry classifications such as the GICS, BICS, ICB, NAICS, SIC, etc.

\item \cite{chen2016industry,kruger2012categorization} describe a behavioral bias: investors overly rely on the standard industry classifications (e.g., SIC, NAICS). Corporate managers can exploit this behavioral bias to steer their company towards a more favorable industry and therefore benefit from a lower cost of capital. The article \cite{chen2016industry} does not discuss this behavioral bias from the investor point of view, but it can pay to have a different (statistical) view of the standard industry classification to avoid such opportunistic window dressing. In \cite{kruger2012categorization}, authors claim that long-short strategies exploiting mispricing due to the industry categorization bias generate statistically significant and economically sizable risk-adjusted excess returns.

\item In \cite{hoberg2017text}, authors show that considering alternative industry classifications (for example, text-based ones) can enhance the returns of well-known quantitative strategies such as the industry momentum one which is driven, according to the paper, by inattention to less visible horizon peers.

\item The same authors find in \cite{hoberg2012stock} that their text-based classification of peers significantly outperform traditional industry peers in explaining firm valuations and peer comovement in the stock market: 
Firms with peers whose products more closely
match theirs have higher stock-market comovement; 
Firms that have more unique products relative to their peers have higher stock market valuations.

\item In \cite{wu2015centrality}, the author investigates the supply chain network using several centrality measures. He finds that ``the
stock performance of supplier central portfolios tends to predict the movements of the overall stock market.'' 
More generally, the author suggests that ``shocks to the supply chain can generate ripple effects, which can show up potentially as lead-lag predictive relations in security returns, earning surprises, default probabilities, and/or distinctive term structure patterns in realized and option implied volatilities and credit default swaps.''

\item Authors of \cite{lopez2018detection} suggests using hierarchical clustering to estimate two parameters required by their ``False Strategy'' theorem, namely the number K of effectively independent tests, and the variance of the Sharpe ratios across the K effectively independent tests. This number K corresponds to the number of clusters of strategies found considering a distance based on the correlation of their returns.
Using these estimates, they can conclude how likely a strategy is spurious.

\item In \cite{cordoba2018anticipating}, authors monitor the correlation network statistics of the constituents of a stock index, and find that an abrupt change in the network is predictive of significant falls in the price of this index. \cite{spelta2017financial} aims at predicting such abrupt network changes.

\end{itemize}


\subsection{Risk Management}

How much money a given portfolio can lose? 
in normal market conditions?
in stressed market conditions?
in the presence of systemic risk?

To answer these questions, the use of clusters and networks can help.
As presented previously, the clustering hierarchy can be used to filter a correlation \cite{tola2008cluster,tumminello2008shrinkage} or a tail dependence \cite{durante2015portfolio} matrix, which helps to measure the risk in normal and stressed market conditions respectively.
The systemic risk as defined by the Bank for International Settlements
is the risk that a failure of a participant to meet its contractual obligations may in turn cause other
participants to default, with the chain reaction leading to broader financial difficulties.
Networks seem thus a particularly relevant tool to study this kind of risk.

\begin{itemize}

\item Study of \textbf{systemic risk}:
\begin{itemize}

\item In \cite{harmon2010networks}, authors assert that the diminution of regulation has removed barriers between sectors and regions allowing bank to diversify their risk, 
but it also increased the economic risk through increased interdependencies.


\item The paper \cite{lautier2011systemic} is focused on energy derivative markets, and their market integration which can be seen as a necessary condition for the propagation of price shocks. 
The MST is used to ``identify the most probable and the shortest path for the transmission of price shocks''. 

\item Authors in \cite{bhattacharjee2017network,bhattacharjee2017investigating} identify the set of indices possessing influence over the Asian region, namely the Hong Kong, Singapore and South Korea ones. Authors claim that the findings of their study can be utilized in effective systemic risk management and for the selection of an optimally-diversified portfolio, resilient to system-level shocks.




\item Authors in \cite{meng2014systemic} focus on the US housing market.
According to the paper, ``dramatic increases in the systemic risk are usually accompanied by regime shifts, which provide a means of early detection of housing bubbles.'' 
They find a sharp increase in housing market correlations over the past decade, indicating that systemic market risk has also greatly increased; They observe that prices diffuse in complex ways that do not require geographical clusters unlike worldwide stock markets which exhibit clear geographical clustering \cite{song2011evolution}.


\item The paper \cite{zhang2014systemic} is focused on the shipping market.
Authors explore the connections between the shipping market and the financial market: 
The shipping market can provide efficient warning before market downturn.
Alike many economic systems which have been exhibiting an
increase in the correlation between different market sectors, a factor that exacerbates
the level of systemic risk, the three major world shipping
markets, (i) the new ship market, (ii) the second-hand ship market, and (iii) the freight market,
have experienced such an increase. Authors show it using the MST, Granger causality analysis, and Brownian distance on the prices of the real shipping market, and the stock
prices of publicly-listed shipping companies.


\item \cite{billio2012econometric} investigates the monthly returns of hedge funds, banks, broker/dealers, and insurance companies. They find that all four sectors have become highly interrelated over the past decade, likely increasing the level of systemic risk.


\item \cite{sandhu2015market} shows that Ricci curvature may serve as an indicator of fragility in the context of financial networks.


\item \cite{papenbrock2015handling} detects distinct correlation
regimes between 1998 and 2013. These correlation regimes have been significantly different
since the financial crisis of 2008 than they had been previously.
Cluster tracking shows that asset classes are now less separated.
Correlation networks help the authors to identify ``risk-on'' and ``risk-off'' assets.

\item For authors in \cite{munnix2012identifying}, the identification of market states based on correlation matrices can be helpful to build an ``early warning system'' for financial markets. This system could be implemented by comparing the current state to previous similar states or monitoring rapid changes in the correlation structure.


\item In \cite{musmeci2015risk}, authors study the clusters' composition evolution, and their persistence.
They observe that the clustering structure is quite stable in
the early 2000s becoming gradually less persistent before the unfolding
of the 2007-2008 crisis. 
The correlation structure eventually recovers persistence
in the aftermath of the crisis, settling up a new phase which is distinct
from the pre-crisis structure one, where the market structure is less related
to industrial sector activity.


\item \cite{huang2016financial} finds that financial institutions which have, in the correlation networks, greater node strength, larger node betweenness centrality, larger node closeness centrality and larger node clustering coefficient tend to be associated with larger systemic risk contributions.

\item \cite{squartini2013early,battiston2016complexity} discuss the detection of early-warning signals of the 2008 crisis via the analysis of the properties of interbank networks.

\item Authors in \cite{almog2019structural} define the structural entropy of a network, which is simply the entropy of the probability vector encoding the proportional size of the clusters in the network. They propose to use structural entropy to monitor the structure of correlation-based networks over time. They suggest this quantity could be an early warning indicator of financial crises.
They observe ``a remarkably high linear correlation between the new measure and the volatility of the assets’ prices over time''.

\item In \cite{cimini2015systemic}, authors highlight that the underlying financial network required to study systemic risk is only partially observable in general. They propose a method to reconstruct such a network, i.e. to build a set of (directed and weighted) dependencies among the constituents of a complex system.

\end{itemize}

\item Risk management \textbf{methods}:

\begin{itemize}

\item In \cite{durante2014clustering}, authors 
design clusters that tend to be comonotonic in their extreme low values: 
To avoid contagion in the portfolio during risky scenarios,
an investor should diversify over these clusters.


\item As far
as diversification is concerned, portfolio managers should probably focus on the most stable parts of the graph \cite{lautier2011systemic}.


\item In \cite{morales2014dependency}, authors postulate the
existence of a hierarchical structure of risks which can be deemed responsible for both stock multivariate dependency structure and univariate multifractal behaviour, and then propose a model that reproduces the empirical observations (entanglement of univariate multi-scaling and multivariate
cross-correlation properties of financial time series).
The interplay between multi-scaling and average cross-correlation is confirmed in \cite{buonocore2018interplay}.


\item Industries (e.g. clusters as statistical industry classification) can be used as risk factors in multifactor risk models \cite{kakushadze2016statistical}.

\item Clusters (statistical industry classification) can be an alternative to sometimes unavailable ``fundamental'' industry classifications (e.g. in emerging or small markets) \cite{kakushadze2016statistical}.

\item In \cite{barfuss2016parsimonious}, authors apply the TMFG for building sparse forecasting models and for financial applications such as stress-testing and risk allocation.

\item \cite{letizia2018corporate} predicts credit risk based on local properties of the network of payments between firms.



\end{itemize}

\end{itemize}

We found that the risk literature using correlation networks and clusters consists essentially in descriptive studies. For now, there are only too few propositions in the academic literature to build effective network-based or cluster-based risk systems.

\subsection{Financial Policy Making}

Clusters and networks can help designing financial policies.
Several papers propose to leverage them to detect risky market environments, 
develop indicators that can predict forthcoming crisis or economic recovery \cite{zhang2011will}, improve economic nowcasting \cite{elshendy2017big},
or find key markets and assets that drive a whole region, and on which stimulus can be applied effectively.

\begin{itemize}

\item Authors of \cite{harmon2010networks} claim that ``separation prevents failure propagation and connections increase risks of global crises'' whereas the prevailing view in favor of  
deregulation is that banks, by investing in diverse sectors,
would have greater stability.
To support their argument, using financial networks, they study the aftermath of the Glass-Steagall Act (1933) repeal by Clinton administration in 1999.
They find that erosion of the Glass–Steagall Act, and cross sector investments eliminated ``firewalls'' that could have prevented the housing sector decline from triggering a wider financial and economic crisis:
\begin{quote}
Our analysis implies that the investment across economic
sectors itself creates increased cross-linking of otherwise much more weakly coupled parts of
the economy, causing dependencies that increase, rather than decrease, risk.
\end{quote}

\item According to \cite{billio2012econometric}, bank and insurance capital requirements and risk management practices based on VaR, which are intended to ensure the soundness of individual financial institutions, may amplify aggregate fluctuations if they are widely adopted:
\begin{quote}
For example, if the riskiness of assets held by one bank increases due to heightened market volatility, to meet its VaR requirements the bank will have to sell some of these risky assets. This liquidation may restore the bank's financial soundness, but if all banks engage in such liquidations at the same time, a devastating positive feedback loop may be generated unintentionally. 
These endogenous feedback effects can have significant implications for the returns of financial institutions, including autocorrelation, increased correlation, changes in volatility, Granger causality, and, ultimately, increased systemic risk, as our empirical results seem to imply.
\end{quote}

\item In \cite{lautier2011systemic}, authors find that the move towards integration started some time ago and there is probably no way to stop or refrain it. 
However, regulation authorities may act in order to prevent prices shocks
from occurring, especially in places where their impact may be important.

\item In \cite{denev2015probabilistic}, the author shows how Bayesian networks (and probabilistic graphical models in general) can be used to model complex networks of debt dependencies between firms. These particular types of networks allow for causal and counterfactual reasoning: What would happen if a set of institutions defaults? What would happen if this institution is bailed out?

\end{itemize}

\section{Practical Fruits of Clusters, Networks, and Hierarchies\protect\footnote{reference to the book \textit{Practical Fruits of Econophysics} \cite{takayasu2006practical}}}

\subsection{Stylized facts}

Stylized facts can be described as follows \cite{cont2001empirical}:
\begin{quote}
A set of [statistical] properties, common across many instruments, markets and time periods, [which] has been observed by independent studies.
\end{quote}

From the papers we reviewed, we can list the following stylized facts:

\begin{itemize}[noitemsep]
\item Firms belonging to some economic sectors are strongly connected within themselves, whereas others are much less connected.
\item The Energy and Financial sectors are examples of strong connections whereas elements belonging to the Conglomerates, Consumer cyclical, Transportation, and Capital Goods sectors are weakly connected.
\item General Electric is at the center of US stocks networks (for several centrality criteria) \cite{mantegna1999hierarchical,bonanno2001high,onnela2003dynamics,brida2008multidimensional}.
\item The Energy, Technology, and Basic Materials sectors are sectors of elements significantly connected among them but weakly interacting with stocks belonging to different economic sectors.
\item The Financial sector is strongly connected within, but also to others.
\item The assets of the classic Markowitz portfolio are always located on the outer leaves of the tree \cite{onnela2003dynamics,pozzi2013spread,peralta2016network}. 
\item The maximum eigenvalue of the correlation matrix, which carries most of the correlations, is very large during market crashes \cite{drozdz2000dynamics} (increased value of the mean correlation).
\item The MST shrinks during market crashes \cite{onnela2003dynamics} and contains a low number of clusters \cite{matesanz2015sovereign}.
\item The MST provides a taxonomy which is well compatible with the sector classification provided by an outside institution \cite{mantegna1999introduction,onnela2003dynamics}.
\item Scale free structure of the MST (i.e. the degree of vertices is power law distributed $f(n) \sim n^{-\alpha}$) \cite{vandewalle2001non,kim2002scale,onnela2003dynamics,bonanno2003topology}, but the scaling exponent depends on market period and window width \cite{onnela2002dynamic}.
\item The MST obtained with the one-factor model is very different from the one obtained using real data \cite{bonanno2003topology}. This invalidates the Capital Asset Pricing Model which is based on the one-factor model $r_i(t) = \alpha_i + \beta_i r_M(t) + \epsilon_i(t)$.
\item Stocks compose a hierarchical system progressively structuring as the sampling time horizon increases \cite{tumminello2007correlation,borghesi2007emergence}.
\item The correlation among market indices presents both a fast and a slow dynamics. The slow dynamics is a gradual growth associated with the development and consolidation of globalization. The fast dynamics is associated with events that originate in a specific part of the world and rapidly (in less than 3 months) affect the global system \cite{song2011evolution,meng2014systemic}.
\item Removing the dynamics of the center of mass decreases the level of correlations, but also makes the cluster structure more evident \cite{borghesi2007emergence}.
\item Scale invariance of correlation structure (by subtraction of the market mode) might have important implications for risk management, because it suggests that correlations on short time scales might be used as a proxy for correlations on longer time-horizons \cite{borghesi2007emergence}.
\item The MST is star-like in low-volatility segments, and chain-like in high-volatility segments \cite{zhang2011will}.
\item Volatility shocks always start at the fringe and propagate inwards \cite{zhang2011will}.
\item The ``post-subprime'' regime correlation matrix shows markedly higher absolute correlations than the others \cite{papenbrock2015handling}.
\item In \cite{papenbrock2015handling}, authors find far less asset class separation in the post-subprime period.
\item One can distinguish three types of topological configurations for the companies: (i) important nodes, (ii) links and (iii) dangling ends \cite{vandewalle2001non}.
\item A node keeps the majority of its neighbours. The non-randomness of the stock market topology is thus a robust property \cite{vandewalle2001non}.
\item The largest eigenvector of the correlation matrix is strongly non-Gaussian, tending to uniform - suggesting that all companies participate. Authors find indeed that all components participate approximately equally to the largest eigenvector. This implies that every company is connected with every other company. In the stock market problem, this eigenvector conveys the fact that the whole market ``moves'' together and indicates the presence of correlations that pervade the entire system \cite{plerou2000random}.
\item The measure of the average length of shortest path in the PMFG shows a small world effect present in the networks at any time horizon \cite{tumminello2007correlation}.
\item  Among the 100 largest market capitalization stocks in the NYSE, the auto and lagged intraday correlations play a much more prominent role in 2011-2013 than in 2001-2003 \cite{curme2015lead}.
\item Authors in \cite{curme2015lead} find striking periodicities in the validated lagged correlations, characterized by surges in network connectivity at the end of the trading day.
\item At short time scales, measured synchronous correlations among stock returns tend to be lower in magnitude \cite{epps1979comovements}, but lagged correlations among assets may become non-negligible \cite{toth2009accurate,curme2015emergence}.
\item Banks may be of more concern than hedge funds from the perspective of connectedness \cite{billio2012econometric}.
\item A lack of distinct sector identity in emerging markets \cite{pan2007collective,nguyen2013one}; Few largest eigenvalues deviate from the bulk of the spectrum predicted by RMT (far fewer than for the NYSE) \cite{pan2007collective,nguyen2018analysis}.
\item Emergence of an internal structure comprising multiple groups of strongly coupled components is a signature of market development \cite{pan2007collective}.
\end{itemize}

\subsection{Moot points and controversies}

Though most of the conclusions of empirical studies do agree, we find some claims that seem to be contradictory:

\begin{itemize}[noitemsep]
\item \cite{kaya2014eccentricity} finds that eccentricity-based risk budgeting portfolios have improved return to risk ratios, hence better invest in centrality (of the minimum spanning tree). On the contrary, \cite{onnela2003dynamics,pozzi2013spread,peralta2016network} conclude that it is better to invest in the peripheries (of the minimum spanning tree).
\item Volatility shocks always start at the fringe and propagate inwards \cite{zhang2011will}, but in \cite{smith2009spread}, authors assert that the credit crisis spreads among affected stocks from more centralized to more outer ones, as spread the news about the extent of damage to the global economy.
\item One might expect that the higher the correlation associated to a link in a correlation-based network is, the higher the reliability of the link is. The paper \cite{tumminello2007spanning} shows that it is not always observed empirically. However, the Cramér–Rao lower bound (CRLB) for correlation \cite{DBLP:conf/ssp/MartiAND16} points out that the higher the correlation, the easier its estimation, i.e. less statistical uncertainty for high correlations.
\item For filtering the correlation matrix, SLCA is more stable than ALCA according to \cite{tumminello2010correlation}, but ALCA is more stable and appropriate than SLCA according to \cite{tola2008cluster,papenbrock2015handling}.
\item During a crisis period, is there an increase or decrease of clusters stability? Most papers find a decrease (e.g. \cite{lautier2011systemic,musmeci2015risk}), but at least one \cite{kocheturov2014dynamics} (using an alternative clustering methodology, the p-median problem) advocates for an increase.
\item In many papers, researchers observe that Minimum Variance portfolios tend to select the same assets as Network-based portfolios \cite{onnela2003dynamics,peralta2016network,pozzi2013spread}; In \cite{huttner2018portfolio}, authors show that in general there are no relations between the two methods, suggesting any empirical relations observed come from special properties of the financial correlations. Which ones? This question is left open to these days.
\end{itemize}

\section{Conclusion}

Correlation networks (and complex networks in general) provide a useful set of quantitative tools for several tasks in finance, e.g. monitoring systemic risk and acting preemptively on a few institutions to prevent cascading effects; building diversified portfolios of assets or strategies; designing statistical arbitrage strategies; residualizing returns by removing the main hierarchical risk factors; finding less crowded cross-sectional risk premia.

However, in our opinion, researchers in this field need to overcome several challenges so that results and techniques exposed in this review can become part of the standard toolbox of the practitioner. We can observe that there is i) a lack of reproducibility for a couple of studies (hence some contradictory claims); ii) difficulty to compare methods due to re-implementation biases and lack of open source code; and iii) no widely accepted common tasks or benchmarks; iv) data is not shared and most of the time confidential.

To tackle these issues, we recommend researchers to provide code and data (or at least synthetic datasets). When the underlying data is confidential, the use of Generative Adversarial Networks (GANs) or other generative models could be a solution to i) share anonymized synthetic data having properties similar to the original dataset; ii) define a common task and iii) set benchmark results enticing other researchers to challenge. This approach has benefited the progress of machine learning as a research field. We think it should help researchers in correlation networks (and other complex networks) having more impact as well.


\section*{Acknowledgments}

Many thanks to the researchers which have contributed to improve this review (in chronological order): David Matesanz, Tiziana Di Matteo, Diego Garlaschelli, Damiano Brigo, Fabrizio Lillo, Quang Nguyen, Thomas Guhr.

\pagebreak

\begin{tiny}
\bibliographystyle{model1-num-names}
\bibliography{sample.bib}
\end{tiny}

\end{document}